\documentstyle[12pt,graphicx]{article}

\begin{document}
\title{Self-avoiding polygons on the square lattice}
\author{Iwan Jensen and Anthony J Guttmann  \\
Department of Mathematics \& Statistics, The University of Melbourne\\
Parkville, Vic. 3052, Australia}
\date{\today}
\maketitle
\bibliographystyle{plain}
\begin{abstract}
We have developed an improved algorithm that allows us to enumerate the number
of self-avoiding polygons on the square lattice to perimeter length 90. 
Analysis of the resulting series yields very accurate estimates of the 
connective constant $\mu =2.63815852927(1)$ (biased) and the critical exponent
$\alpha =  0.5000005(10)$ (unbiased). The critical point is indistinguishable
from a root of the polynomial $581x^4 + 7x^2 - 13 =0.$ An asymptotic
expansion for the coefficients is given for all $n.$ There is strong evidence
for the absence of any non-analytic correction-to-scaling exponent.
\end{abstract}

\section{Introduction}

A self-avoiding polygon (SAP) can be defined as a walk on a lattice
which returns to the origin and has no other self-intersections.
The history and significance of this problem is nicely discussed
in \cite{Hug}. Alternatively we can define a SAP as a connected 
sub-graph (of a lattice) whose vertices are of degree 0 or 2. 
Generally SAPs are considered distinct up to a translation, so if there 
are $p_n$ SAPs of length $n$ there are $2np_n$ walks (the factor of 
two arising since the walk can go in two directions).  The enumeration of 
self-avoiding polygons on various lattices is an interesting combinatorial 
problem in its own right, and is also of considerable importance in the 
statistical mechanics of lattice models \cite{Hug}. 

The basic problem is the calculation of the generating function

\begin{equation}
P(x)  =  \sum_{n} p_{2n} x^{2n} \sim A(x) + B(x)(1-x^2/x_c^2)^{2-\alpha},
\end{equation}
\noindent
where the functions $A$ and $B$ are believed to be regular in the
vicinity of $x_c.$ We discuss this point further in Sec. 3, as it
pertains to the presence or otherwise of a non-analytic
correction-to-scaling term. Despite strenuous effort over the past 50 years 
or so this problem has not been solved on any regular two dimensional lattice.
However, much progress has been made in the  study of various restricted
classes of polygons and many problems have been solved exactly. These 
include staircase polygons \cite{Temperley,Polya,Delest,Brak90,Lin91}, 
convex polygons \cite{Delest,Kim,GE88b,Lin88}, row-convex polygons 
\cite{Brak90,Lin91}, and almost convex polygons \cite{Lin92}. 
Also, for the hexagonal lattice the critical point, $x_c^2=1/(2+\sqrt{2})$ 
as well as the critical exponent $\alpha = 1/2$ are known exactly 
\cite{Nienhuis,Baxter}, though non-rigorously. Very firm evidence exists from
previous numerical work that the exponent $\alpha$ is universal and thus 
equals 1/2 for all two dimensional lattices \cite{GE88a,EG89,JG98}. Thus the 
major remaining problem, short of an exact solution, is the calculation of 
$x_c$ for various lattices. Recently the authors found a simple mapping 
between the generating function for SAPs on the hexagonal lattice and the 
generating function for SAPs on the $(3.12^2)$ lattice \cite{JG98}. Knowledge 
of the exact value for $x_c$ on the hexagonal lattice resulted in the 
exact determination of the critical point on the $(3.12^2)$ lattice.

In order to study this and related systems, when an exact solution
can't be found one has to resort to numerical methods. For many problems
the method of series expansions is by far the most powerful method
of approximation. For other problems Monte Carlo methods are superior.
For the analysis of $P(x)$, series analysis is undoubtedly the
most appropriate choice. This method consists of calculating
the first coefficients in the expansion of the generating function. 
Given such a series, using the numerical technique known as 
differential approximants \cite{Guttmann89}, 
highly accurate estimates can frequently be obtained for  the 
critical point and exponents, as well as the location and critical 
exponents of possible non-physical singularities.

This paper builds on the work of Enting \cite{Ent} who enumerated 
square lattice polygons to 38 steps using the finite
lattice method. Using the same technique this enumeration was extended 
by Enting and Guttmann to 46 steps \cite{EG85} and later to 56 steps 
\cite{GE88a}. Since then they extended the enumeration to 70 steps
in unpublished work. These extensions to the enumeration were largely
made possible by improved computer technology. In this work we have 
improved the algorithm and extended the enumeration to 90 steps while 
using essentially  the same computational resources used to obtain 
polygons to 70 steps.

The difficulty in the enumeration of most interesting lattice problems
is that, computationally, they are of exponential complexity. It
would be a great breakthrough if a polynomial time algorithm could be found,
while a linear time algorithm is, to all intents and purposes, equivalent
to an exact solution.  Initial efforts at computer enumeration
of square lattice polygons were based on direct counting. The
computational complexity was proportional to $\lambda_1^n,$ where
$n$ is the length of the polygon, and $\lambda_1 = 1/x_c \approx 2.638.$
The dramatic improvement achieved \cite{Ent} by the finite lattice method
can be seen from its complexity, which is proportional to $\lambda_2^n,$ 
where $\lambda_2 = 3^\frac{1}{4}  \approx 1.316.$ Our new algorithm, 
described below, has reduced both time and storage requirements by virtue 
of a complexity which is proportional to $\lambda_3^n,$ where
$\lambda_3 \approx 1.20.$ It is worth noting that for simpler 
restricted cases it possible to devise much more efficient algorithms.
For problems such as the enumeration of convex polygons \cite{GE88b} 
and almost convex polygons \cite{Enting92,Moraal} the algorithms are of
polynomial complexity. Other interesting and related problems for which
efficient transfer matrix algorithms can be devised include
Hamiltonian circuits on rectangular strips (or other compact shapes) 
\cite{Kloczkowski} and self-avoiding random walks \cite{Derrida,CG96}.

In the next section we will very briefly review the finite lattice
method for enumerating square lattice polygons and give some
details of the improved algorithm. The results of
the analysis of the series are presented in Section~\ref{sec:analysis}
including a detailed discussion of a conjecture for the exact
critical point.

\section{Enumeration of polygons \label{sec:flm}}

\begin{figure}
\begin{center}
\includegraphics{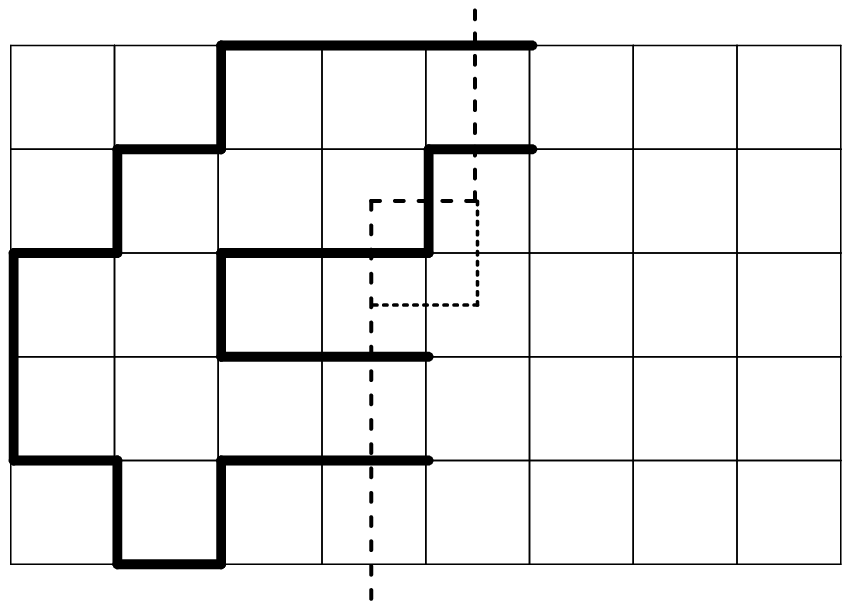}
\end{center}
\caption{\label{fig:transfer}
A snapshot of the intersection (dashed line) during the transfer matrix 
calculation on the square lattice. Polygons are enumerated by successive
moves of the kink in the intersection, as exemplified by the position given 
by the dotted line, so that one vertex at a time is added to the rectangle. 
To the left of the intersection we have drawn an example of a 
partially completed polygon.}
\end{figure}

The method used to enumerate SAP on the square lattice is an
enhancement of the method devised by Enting \cite{Ent} in his 
pioneering work. The first terms in the series for the polygon generating 
function can be calculated using transfer matrix techniques to count 
the number of polygons in rectangles $W+1$ edges wide and $L+1$ edges long. 
The transfer matrix technique  involves drawing a line through the
rectangle intersecting a set of $W+2$ edges. For each configuration of 
occupied or empty edges along the intersection we maintain a (perimeter) 
generating function for loops to the left of the line cutting the 
intersection in that particular pattern. Polygons in a given rectangle 
are enumerated by moving the intersection so as to add one vertex at a time, 
as shown in Fig.~\ref{fig:transfer}. The allowed configurations along the 
intersection are described in \cite{Ent}. Each configuration can be 
represented by an ordered set of edge states $\{n_i\}$, where

\begin{equation}\label{eq:states}
n_i  = \left\{ \begin{array}{rl}
 0 &\;\;\; \mbox{empty edge},  \\ 
1 &\;\;\; \mbox{lower part of loop closed to the left}, \\
2 &\;\;\; \mbox{upper part of loop closed to the left}. \\
\end{array} \right.
\end{equation}
\noindent
Configurations are read from the bottom to the top. So the configuration
along the intersection of the polygon in Fig.~\ref{fig:transfer} is 
$\{0112122\}$. 

The rules for updating the partial generating functions as the
intersection is moved are identical to the original work, so we 
refer the interested reader to \cite{Ent} for 
further details regarding this aspect of the  transfer matrix calculation.

Due to the obvious symmetry of the lattice one need only consider rectangles 
with $L \geq W$. Valid polygons were required to span the enclosing
rectangle in the lengthwise direction. So it is clear that
polygons with projection on the $y$-axis $< W$, that is polygons which
are narrower than the width of the rectangle, are counted many times.
It is however easy to obtain the polygons of width exactly $W$ and
length exactly $L$ from this enumeration \cite{Ent}.
Any polygon spanning such a rectangle has a perimeter
of length at least $2(W+L)$.  By adding the contributions from all 
rectangles of width $W \leq W_{\rm max}$ (where the choice of $W_{\rm max}$ 
depends on available computational resources, as discussed below) and length 
$W \leq L \leq 2W_{\rm max}-W+1$, with contributions from rectangles 
with $L>W$ counted twice, the number of polygons per vertex of an infinite 
lattice is obtained correctly up to perimeter $4W_{\rm max}+2$.

The major improvement of the method used to enumerate polygons in this 
paper is that we require valid polygons to span the rectangle in 
{\em both} directions. In other words we directly enumerate polygons of
width exactly $W$ and length $L$ rather than polygons of width $\leq W$
and length $L$ as was done originally. The only drawback of this approach 
is that for most configurations we have to use four distinct generating
functions since the partially completed polygon  could have reached neither, 
both, the lower, or the upper boundaries of the rectangle. The major 
advantage is that the memory requirement of the algorithm is 
exponentially smaller.  

Realizing the full savings in memory usage requires two
enhancements to the original algorithm. Firstly, for each
configuration we must keep track of the current minimum number of steps 
$N_{\rm cur}$ that have been inserted to the left of the intersection 
in order to build up that particular configuration. Secondly, we 
calculate the minimum number of additional steps $N_{\rm add}$ required to 
produce a valid polygon. There are 
three contributions, namely the number of steps required to close the 
polygon, the number of steps needed (if any) to ensure that the 
polygon touches both the lower and upper boundary, and finally the 
number of steps needed (if any) to extend at least $W$ edges in the 
length-wise direction. If the sum 
$N_{\rm cur}+N_{\rm add} > 4W_{\rm max}+2$ we can discard the partial
generating function for that  configuration because it won't make a 
contribution to the polygon count up to the perimeter lengths we are 
trying to obtain. For instance polygons spanning a rectangle with a width 
close to $W_{\rm max}$ have to be almost convex, so very convoluted 
polygons are not possible. Thus configurations with 
many loop ends (non-zero entries) make no contribution at perimeter 
length $\leq 4W_{\rm max}+2$.

\begin{figure}
\begin{center}
\includegraphics{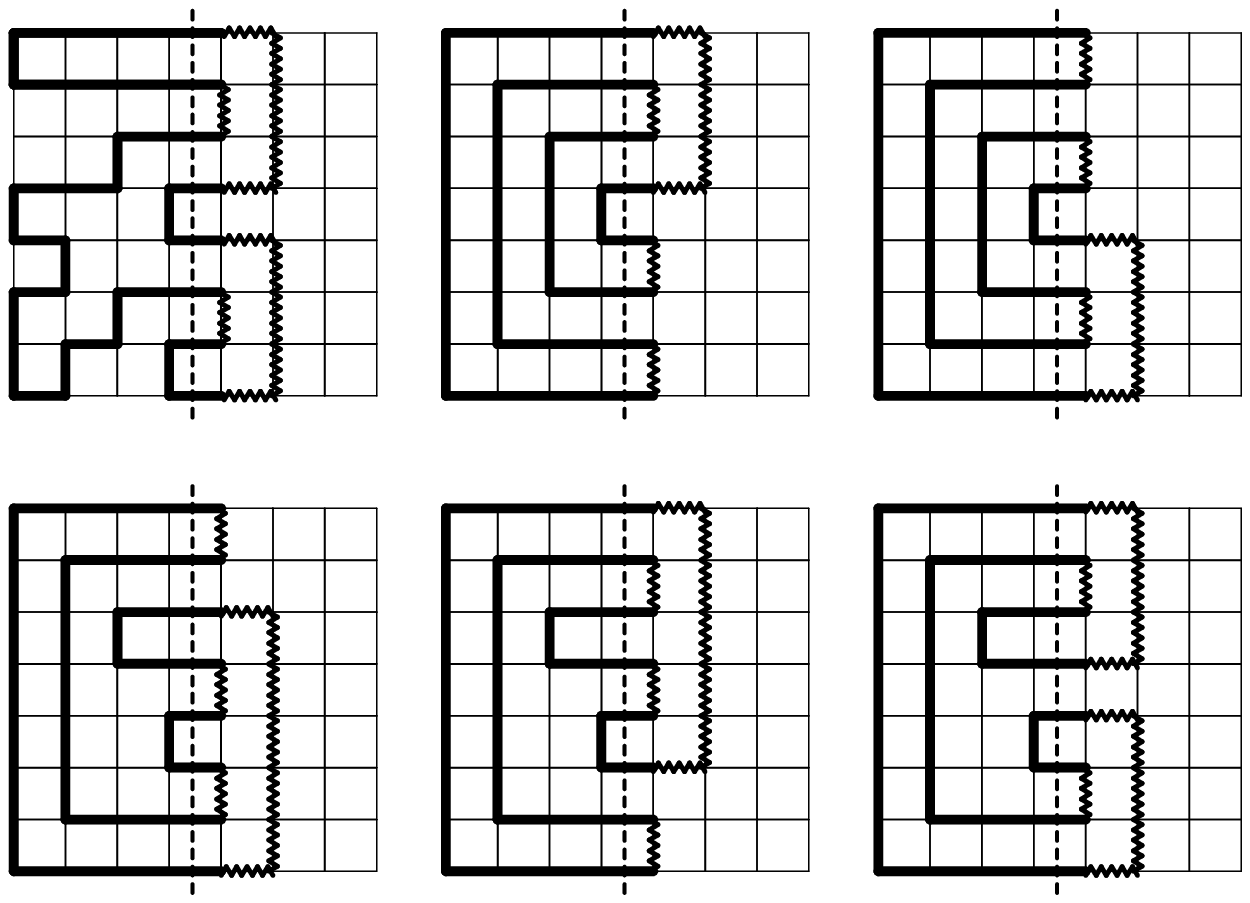}
\end{center}
\caption{\label{fig:polclose} Examples of partially generated
polygons (thick solid lines) to the left of the intersection (dashed line)
and how to close them in a valid way (thick wavy line). Upper left panel 
shows how to close the configuration $\{12112212\}$. The upper middle and
right panels show the two possible closures of the configuration
$\{11112222\}$. The lower panels show the three possible closures of the 
configuration $\{11121222\}$. }
\end{figure}

The number of steps needed to ensure a spanning polygon is straightforward 
to calculate. The complicated part of the new approach is the algorithm 
to calculate the number of steps required to close the polygon. 
There are very many special cases depending on the position of the
kink in the intersection and whether or not the partially completed 
polygon has reached the upper or lower boundary of the bounding rectangle.
So in the following we will only briefly describe some of the simple
contributions to the closing of a polygon. Firstly, if the partial
polygon contains separate pieces these have to be connected as illustrated
in Fig.~\ref{fig:polclose}. Separate pieces are easy to locate since
all we have to do is start at the bottom of the intersection and
moving upwards we count the number of 1's and 2's in the configuration.
Whenever these numbers are equal a separate piece has been found and
(provided one is not at the last edge in the configuration) the
currently encountered 2-edge can be connected to the next 1-edge above.
$N_{\rm add}$ is incremented by the number of steps (the distance) 
between the edges and the two edge-states are removed from the
configuration before further processing. It is a little less obvious
that if the configuration start (end) as $\{112\ldots 2\}$ 
($\{1\ldots 122\}$) the two lower (upper) edges can safely be 
connected (note that there can be any number of 0's interspersed 
before the $\ldots$). Again $N_{\rm add}$ is 
incremented by the number of steps between the edges, and the two 
edge-states are removed from the configuration -- leading to the new
configuration $\{001\ldots 2\}$ ($\{1\ldots 200\}$) -- before 
further processing. After
these operations we may be left with a configuration which has
just one 1- and one 2-edge, in which case we are done since these
two edges can be connected to form a valid polygon.  This is illustrated 
in Fig.~\ref{fig:polclose} where the upper left panel shows how
to close the partial polygon with the intersection $\{12112212\}$, which
contain three separate pieces. After connecting these pieces we are
left with the configuration $\{10012002\}$. We now connect the two 1-edges
and note that the first two-edge is relabeled to a 1-edge (it has
become the new lower end of the loop). Thus we get the configuration 
$\{00001002\}$ and we can now connect the remaining two edges
and end up with a valid completed polygon. Note that in the last two
cases, in addition to the steps spanning the distance between the edges,
an additional two horizontal steps had to be added in order to form a
valid loop around the intervening edges.  If the transformation
above doesn't result in a closed polygon we must have 
a configuration of the form $\{111\ldots 222\}$. The difficulty lies 
in finding the way to close such configurations with the
smallest possible number of additional steps. Suffice to say that
if the number of non-zero entries is small one can easily devise
an algorithm to try all possible valid ways of closing a polygon
and thus find the minimum number of additional steps. 
In Fig.~\ref{fig:polclose} we show all possible ways of closing polygons
with 8 non-zero entries. Note that we have shown the generic cases
here. In actual cases there could be any number of 0-edges interspersed
in the configurations and this would determine which way of closing
would require the least number of additional steps.

\begin{figure}
\begin{center}
\includegraphics{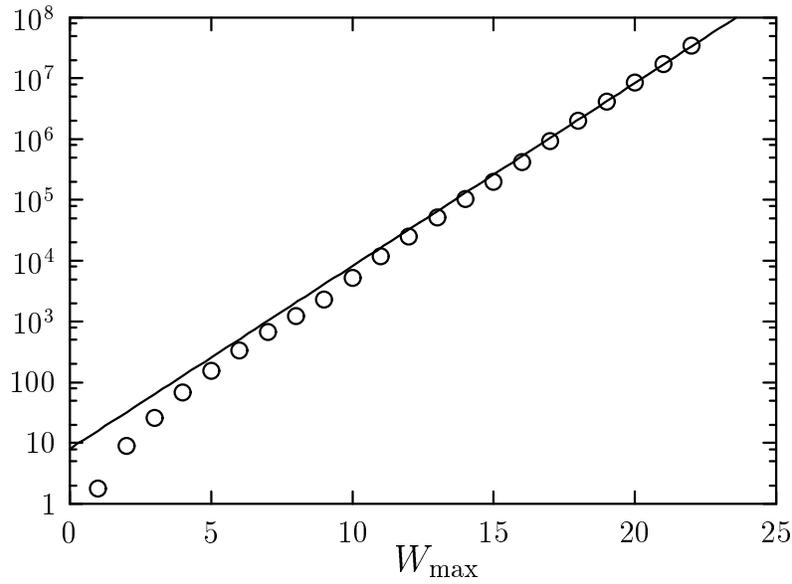}
\end{center}
\caption{\label{fig:numconf}
The number of configurations required as  $W_{\rm max}$ is increased.
The straight line corresponds to a growth factor $\lambda = 2$.}
\end{figure}

With the original algorithm the number of configurations required 
as $W_{\rm max}$ increased grew asymptotically as $3^{W_{\rm max}}$
\cite{GE88a}. Our enumerations indicate that the computational complexity is 
reduced significantly. While the number of configurations still grows
exponentially as $\lambda^{W_{\rm max}}$ the value of $\lambda$
is reduced from $\lambda = 3$ to $\lambda \simeq 2$ with the 
improved algorithm (Fig.~\ref{fig:numconf} shows the number
of configuration required as $W_{\rm max}$ increases). Furthermore, for any 
$W$ we know that contributions will start at $4W$ since the smallest 
polygons have to span a $W\times W$ rectangle. So for each configuration 
we need only retain $4(W_{\rm max}-W)+2$ terms of the generating 
functions while in the original algorithm contributions started at 
$2W$ because the polygons were required to span only in the length-wise 
direction. We also note that on the square lattice all SAP's are of even
length so for each configuration every other term in the generating 
function is zero, which allows us to discard half the terms and retain
only the non-zero ones.

Finally a few remarks of a more technical nature. The number of contributing 
configurations becomes very sparse in the total set of possible states along 
the boundary line and as is standard in such cases one uses a hash-addressing 
scheme \cite{Mehlhorn}. Since the integer coefficients occurring in the series 
expansion become very large, the calculation was performed using modular 
arithmetic \cite{Knuth}. This involves performing the calculation modulo 
various prime numbers $p_i$ and then reconstructing the full integer
coefficients at the end. In order to save memory we used primes of the form 
$p_i=2^{15}-r_i$ so that the residues of the coefficients in the polynomials 
could be stored using 16 bit integers. The Chinese remainder theorem
ensures that any integer has a unique representation in terms of residues. If 
the largest absolute values occurring in the final expansion is $m$, then we 
have to use a number of primes $k$ such that $p_1p_2\cdots p_k/2 > m$.  Up to 
8 primes were needed to represent the coefficients correctly. 

Combining all the memory minimization tricks mentioned above allows us to 
extend the series for the square lattice polygon generating function from 
70 terms to 90 terms using at most 2Gb of memory. Obtaining 
a series this long with the original algorithm would have required
at least 200 times as much memory, or close to half a terabyte!
The calculations were performed on an 8 node AlphaServer 8400 with a
total of 8Gb memory. The total CPU time required was about a week
per prime. Obviously the calculation for each width and prime are
totally independent and several calculations were done simultaneously.

In Table~\ref{tab:series} we have listed the new terms obtained
in this work. They of course agree with the terms up to length 70 
computed using the old algorithm. The number of polygons of length 
$\leq 56$ can be found in \cite{GE88a}.

\section{Analysis of the series \label{sec:analysis}}

We analyzed the series for the polygon generating function by the
numerical method of differential approximants \cite{Guttmann89}.
In Table~\ref{tab:analysis}  we have listed estimates for the
critical point $x_c^2$  and exponent $2-\alpha$
of the series for the square lattice SAP generating function.  
The estimates were obtained by averaging values obtained from first order
$[L/N;M]$ and second order $[L/N;M;K]$ inhomogeneous differential
approximants. For each order $L$ of the inhomogeneous polynomial we averaged 
over those approximants to the series which used  at least the first
35 terms of the series (that is, polygons of perimeter at least 74),
and used approximants such that the difference between $N$, $M$, 
and $K$ didn't exceed 2. These are therefore ``diagonal'' approximants.
Some approximants were excluded from 
the averages because the estimates were obviously spurious. The error
quoted for these estimates reflects the spread (basically one standard
deviation) among the approximants. Note that these error bounds should
{\em not} be viewed as a measure of the true error as they cannot include
possible systematic sources of error. We discuss further the systematic 
error when we consider biased approximants. Based on these estimates we 
conclude that $x_c^2 = 0.1436806289(5)$ and $\alpha = 0.5000005(10)$.

\begin{figure}
\begin{center}
\includegraphics{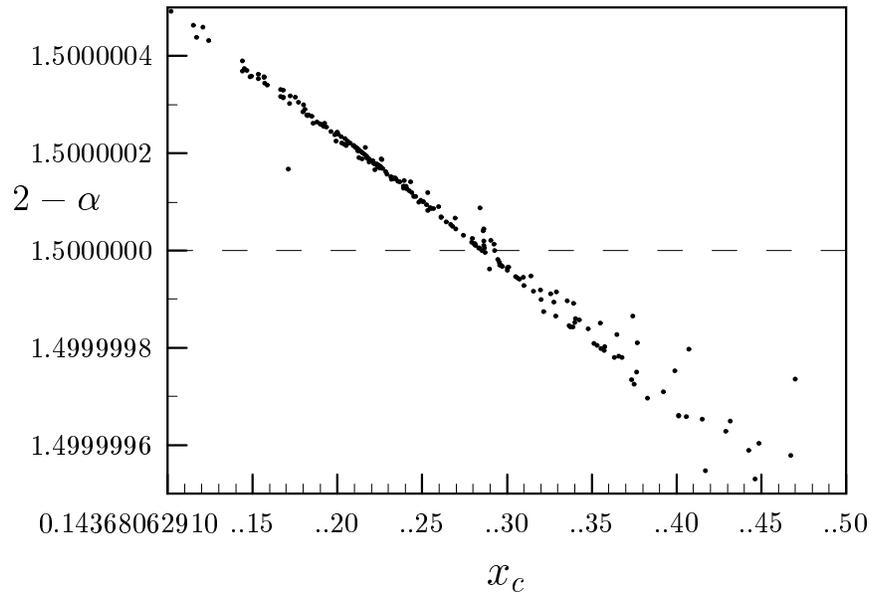}
\end{center}
\caption{\label{fig:crpexp} Estimates for the critical exponent
$2-\alpha$ vs. estimates for the critical point $x_c^2$ of the
square lattice polygon generating function.}
\end{figure}

As stated earlier there is very convincing evidence that the
critical exponent $\alpha = 1/2$ exactly. If we assume this to
be true we can obtain a refined estimate for the critical point
$x_c^2$. In Fig.~\ref{fig:crpexp} we have plotted estimates
for the critical exponent $2-\alpha$ against estimates for the 
critical point $x_c^2$. Each dot in this figure represents a
pair of estimates obtained from a second order inhomogeneous 
differential approximant. The order of the inhomogeneous polynomial
was varied from 0 to 10. We observe that there is an almost
linear relationship between the estimates for $2-\alpha$ and
$x_c^2$ and that for $2-\alpha=3/2$ we get 
$x_c^2\simeq  0.14368062928\ldots$. In order to get some idea as to the 
effect of systematic errors, we carried out this analysis using polygons 
of length up to 60 steps, then 70, then 80 and finally 90 steps. The 
results were $x_c^2 = 0.1436806308$ for $n=60,$ $x_c^2 = 0.14368062956$ 
for $n=70,$ $x_c^2 = 0.14368062930$ for $n=80,$ and $x_c^2 = 0.14368062928$ 
for $n=90.$ This is a rapidly converging sequence of estimates, though we 
have no theoretical basis that would enable us to assume any particular 
rate of convergence. However, observing that the differences between 
successive estimates are decreasing by a factor of at least 5, it is not 
unreasonably optimistic to estimate the limit at $x_c^2 = 0.14368062927(1).$

This leads to our final estimate $x_c^2=  0.14368062927(1)$ and thus we find 
the connective constant $\mu =1/x_c=2.63815853034(10)$. It is interesting to 
note that some years ago we \cite{CG92} pointed out that since the hexagonal 
lattice connective constant is given by the zero of a quadratic in $x^2,$ it 
is plausible that this might be the case also for the square lattice 
connective constant. On the basis of an estimate of the connective constant 
that was 4 orders of magnitude less precise, we pointed out then that the 
polynomial $$581x^4 + 7x^2 -13 =0$$ was the only polynomial we could find 
with ``small'' integer coefficients consistent with our estimate. The 
relevant zero of this polynomial is $x_c^2=0.1436806292698685..$ 
in complete agreement with our new estimate --- which, as noted above, 
contains four more significant digits! Unfortunately the other zero is at 
$x_c^2 = -0.1557288\ldots,$ and we see no evidence of such a singularity.
Nevertheless, the agreement is so astonishingly good that we are
happy to take this as a good algebraic approximation to the connective
constant. An argument as to why we might not expect to see the singularity 
on the negative real axis from our series analysis would make the root of 
the above polynomial a plausible conjecture for the exact value, but at 
present such an argument is missing.

Two further analyses were carried out on the data. Firstly, a study
of the location of non-physical singularities, and secondly, a study
of the asymptotic form of the coefficients --- which is relevant to
the identification of any correction-to-scaling exponent. 
Singularities outside the radius of convergence give exponentially
small contributions to the asymptotic form of the coefficients, so
are notoriously hard to analyse. Nevertheless, we see clear evidence
of a singularity on the negative real axis at $x^2 \approx -0.40$
with an exponent that is extremely difficult to analyse but could
be $1.5,$ in agreement with the physical exponent. There is weaker
evidence of a further conjugate pair of singularities. First order
approximants locates these at $-0.015 \pm 0.36i,$ while second
order approximants locates them at $-0.035 \pm 0.31i.$ There is
also evidence of a further singularity on the negative real axis
at $x_c^2 = -0.7.$ We are unable to give a useful estimate of the 
exponents of these singularities.

We turn now to the asymptotic form of the coefficients. We have argued
previously \cite{CG96} that there is no non-analytic correction-to-scaling
exponent for the polygon generating function. This is entirely consistent
with Nienhuis's \cite{Nienhuis} observation that there is a 
correction-to-scaling exponent of $\Delta = \frac{3}{2}.$ Since for the 
polygon generating function exponent $\alpha = \frac{1}{2},$ the correction 
term has an exponent equal to a positive integer, and therefore 
``folds into'' the analytic background term, denoted $A(x)$ in Eqn.(1).
This is explained in greater detail in \cite{CG96}.
We assert that the asymptotic form for the polygon generating function
is as given by Eqn.(1) above. In evidence of this, we remark
that from (1) follows the asymptotic form
\begin{equation}
p_{2n}x_c^{2n} \sim n^{-\frac{5}{2}}[a_1 + a_2/n + a_3/n^2 
                   + a_4/n^3 + \cdots].
\end{equation}
Using our algebraic approximation to $x_c$ quoted above, we show in
Table~\ref{tab:fit} the estimates of the amplitudes $a_1, \cdots, a_4.$ From 
the table we see that $a_1 \approx 0.0994018,$ $a_2 \approx -0.02751,$ 
$a_3 \approx 0.0255$ and $a_4 \approx 0.12,$ where in all cases we expect the 
error to be confined to the last quoted digit. The excellent convergence of 
all columns is strong evidence that the assumed asymptotic form is correct. 
If we were missing a term corresponding to, say, a half-integer correction, 
the fit would be far worse. This is explained at greater length in \cite{CG96}.
So good is the fit to the data that if we take the last entry in the table, 
corresponding to $n = 45,$ and use the  entries as the amplitudes, then 
$p_4 \cdots p_{16}$ are given exactly by the above asymptotic form (provided 
we round to the nearest integer), and beyond perimeter $20$ all coefficients 
are given to the same accuracy as the leading amplitude.

Finally, to complete our analysis, we estimate the critical amplitudes
$A(x_c^2)$ and $B(x_c^2),$ defined in Eqn.(1). 
$A(x_c^2)$ has been estimated by evaluating Pad\'e approximants to
the generating function, evaluated at $x_c^2.$
In this way we estimate $A(x_c^2) \approx 0.036,$
while $B(x_c^2)$ follows from the estimate of $a_1$ in Eqn.(3), since
$B(x_c^2) = \frac{4\sqrt{\pi}a_1}{3} \approx 0.234913.$

\section{Conclusion}

We have presented an improved algorithm for the enumeration of self-avoiding
polygons on the square lattice. The computational complexity of the algorithm
is estimated to be $1.2^n.$ Implementing this algorithm has enabled us to
obtain polygons up to perimeter length 90. Decomposing the coefficients into
prime factors reveals frequent occurrence of very large prime factors,
supporting the widely held view that there is no ``simple'' formula for
the coefficients. For example, $p_{78}$ contains the prime factor
7789597345683901619. Our extended series enables us to give an extremely
precise estimate of the connective constant, and we give a simple algebraic
approximation that agrees precisely with our numerical estimate.
An alternative analysis provides very strong evidence for the absence
of any non-analytic correction terms to the proposed asymptotic form
for the generating function. Finally we give an asymptotic representation
for the coefficients which we believe to be accurate for all positive
integers.

\section{Acknowledgments}

We gratefully acknowledge valuable discussions with Ian Enting, useful
comments on the manuscript by Alan Sokal and financial 
support from the Australian Research Council.

\newpage

\clearpage

\begin{table}
\caption{\label{tab:series} The number, $x_n$, of embeddings of 
$n$-step polygons on the square lattice. Only non-zero terms are listed.}
\begin{center}
\begin{tabular}{rrrr} \hline \hline
$n$ & $x_n$ & $n$ & $x_n$ \\ \hline 
58 & 59270905595010696944 &  76  & 1158018792676190545425711414 \\
60 & 379108737793289505364 &  78  & 7554259214694896127239818088 \\
62 & 2431560774079622817356 &  80  &  49360379260931646965916677280 \\
64 & 15636142410456687798584 &  82  &  323028185951187646733521902740 \\
66 & 100792521026456246096640 &  84  & 2117118644744425875029583096670 \\
68 & 651206027727607425003232 &  86  &  13895130612692826326409919713700 \\
70 & 4216407618470423070733556 &  88  & 91319729650588816198004801698400 \\ 
72 & 27355731801639756123505014 &  90  & 600931442757555468862970353941700 \\
74 & 177822806050324126648352460 & & \\
\hline \hline
\end{tabular}
\end{center}
\end{table}

\begin{table}
\caption{\label{tab:analysis} Estimates for the critical point
$x_c^2$ and exponent $2-\alpha$ obtained from first and second order
inhomogeneous differential approximants to the series for square lattice
polygon generating function. $L$ is the order of the inhomogeneous
polynomial.}
\begin{center}
\begin{tabular}{lllll} \hline \hline
 $L$   &  \multicolumn{2}{c}{First order DA} & 
       \multicolumn{2}{c}{Second order DA} \\ \hline 
    &  \multicolumn{1}{c}{$x_c^2$} & \multicolumn{1}{c}{$2-\alpha$} & 
      \multicolumn{1}{c}{$x_c^2$} & \multicolumn{1}{c}{$2-\alpha$} \\ \hline 
  1  & 0.14368062897(17) & 1.50000074(35) & 
      0.14368062883(45)& 1.50000092(92) \\
 2  & 0.14368062902(14) & 1.50000068(26) & 
      0.14368062943(29)& 1.49999957(80) \\
 3  & 0.14368062878(35) & 1.50000107(71) & 
      0.14368062914(20)& 1.50000034(51) \\
 4  & 0.14368062910(29) & 1.50000038(61) &  
      0.14368062914(16)& 1.50000038(44) \\
 5  & 0.14368062890(43) & 1.50000085(93) & 
      0.14368062911(53)& 1.5000002(12) \\
 6  & 0.14368062863(49) & 1.5000014(10) & 
      0.14368062901(54)& 1.5000005(12) \\
 7  & 0.14368062886(39) & 1.50000094(80) & 
      0.14368062881(52)& 1.5000009(10) \\
 8  & 0.14368062885(64) & 1.5000008(13)& 
      0.143680629210(97)& 1.50000021(24) \\
\hline \hline
\end{tabular}
\end{center}
\end{table}

\begin{table}
\caption{\label{tab:fit} A fit to the asymptotic form 
$p_{2n}x_c^{2n} \sim n^{-\frac{5}{2}}[a_1 + a_2/n + a_3/n^2 + a_4/n^3 
+ \cdots]$
Estimates of the amplitudes $a_1, a_2, a_3, a_4.$}
\begin{center}
\begin{tabular}{|c|cccc|}
\hline \hline
$n$ & $a_1$ & $a_2$ & $a_3$ & $a_4$  \\ \hline
 20 &  0.09940085 & -0.02745705 &  0.02476376 &  0.11822181  \\
 21 & 0.09940118  &-0.02747548  & 0.02511347 &  0.11601107 \\
 22 &   0.09940140 & -0.02748880 &  0.02537979 &  0.11423855 \\
 23 &  0.09940154  &-0.02749767 &  0.02556592 &  0.11293766 \\
 24 &  0.09940164 & -0.02750397 &  0.02570426 &  0.11192457 \\
 25 &  0.09940170 & -0.02750829 &  0.02580364 &  0.11116357 \\
 26 &  0.09940174 & -0.02751137 &  0.02587757 &  0.11057283 \\
 27 &  0.09940177 & -0.02751355 &  0.02593211 &  0.11011880 \\
 28 &  0.09940179 & -0.02751510 &  0.02597236 &  0.10977030\\
 29 &  0.09940180 & -0.02751619 &  0.02600168 &  0.10950667\\
 30 &  0.09940181 & -0.02751694 &  0.02602273 &  0.10931043\\
 31 &  0.09940182 & -0.02751745 &  0.02603734 &  0.10916929\\
 32 &  0.09940182 & -0.02751777 &  0.02604692  & 0.10907354\\
 33 &  0.09940182 & -0.02751795 &  0.02605254 &  0.10901552\\
 34 &  0.09940182 & -0.02751802 &  0.02605500 &  0.10898929\\
 35 &  0.09940182 & -0.02751802 &  0.02605494 &  0.10898993\\
 36 &  0.09940182 & -0.02751796 &  0.02605285 &  0.10901358\\
 37 &  0.09940182 & -0.02751785 &  0.02604913 &  0.10905699\\
 38 &  0.09940182 & -0.02751771 &  0.02604408 &  0.10911757\\
 39 &  0.09940182 & -0.02751755 &  0.02603796 &  0.10919302\\
 40 &  0.09940182 & -0.02751736 &  0.02603097 &  0.10928158\\
 41 &  0.09940182 & -0.02751717 &  0.02602327 &  0.10938160\\
 42 &  0.09940181 & -0.02751696 &  0.02601500 &  0.10949174 \\
 43 &  0.09940181 & -0.02751675 &  0.02600629 &  0.10961079 \\
 44 &  0.09940181 & -0.02751653 &  0.02599720 &  0.10973796 \\
 45 &  0.09940181 & -0.02751631 &  0.02598785 &  0.10987195 \\
\hline \hline
\end{tabular}
\end{center}
\end{table}


\begin{thebibliography}{10}

\bibitem{Baxter} R. J. Baxter, J. Phys. A {\bf 19}, 2821 (1986).

\bibitem{Brak90} R. Brak and A. J. Guttmann, J. Phys. A {\bf 23}, 4581 (1990).

\bibitem{CG92} A. R. Conway and A. J. Guttmann, 
J. Phys. A. {\bf 26}, 1519 (1993).

\bibitem{CG96} A. R. Conway and A. J. Guttmann, 
Phys. Rev. Letts. {\bf 77}, 5284 (1996).

\bibitem{Delest} M. P. Delest and G. Viennot, 
Theor. Comp. Sci. {\bf 34}, 169 (1984).

\bibitem{Derrida} B. Derrida, J. Phys. A {\bf 14}, L5 (1981).

\bibitem{Ent} I. G. Enting, J. Phys. A. {\bf 13}, 3713 (1980).

\bibitem{EG85} I. G. Enting and A. J. Guttmann, 
J. Phys. A. {\bf 18}, 1007 (1985).

\bibitem{EG89} I. G. Enting and A. J. Guttmann,
J. Phys. A. {\bf 22}, 1371 (1989).

\bibitem{Enting92} I. G. Enting, A. J. Guttmann, L. B. Richmond and
N. C. Wormald, Random Structures and Algorithms {\bf 3}, 445 (1992).

\bibitem{GE88a} A. J. Guttmann and I. G. Enting, 
J. Phys. A. {\bf 21}, L165 (1988).

\bibitem{GE88b} A. J. Guttmann and I. G. Enting, 
J. Phys. A. {\bf 21}, L467 (1988).


\bibitem{Guttmann89} A. J. Guttmann, in {\em Phase Transitions and
Critical Phenomena}, Vol. 13, eds. C Domb and J L Lebowitz,
Academic Press, New York (1989).

\bibitem{Hug} B. D. Hughes, in {\em Random walks and random environments,
Vol. I Random walks}, Clarendon Press, Oxford (1995).

\bibitem{JG98} I. Jensen  and A. J. Guttmann,
J. Phys. A. {\bf 31}, 8137 (1998).

\bibitem{Kim} D. Kim, Discrete Math. {\bf 70}, 47 (1988).

\bibitem{Kloczkowski} A. Kloczkowski and R. L. Jernigan,
J. Chem. Phys. {\bf 109}, 5134 and 5147 (1998).

\bibitem{Knuth} D. E. Knuth, {\em Seminumerical Algorithms (The Art of
Computer Programming 2)}, Addison-Wesley, Reading, MA (1969). 

\bibitem{Lin88} K. Y. Lin and S. J. Chang, J. Phys. A. {\bf 21}, 2635 (1988).

\bibitem{Lin91} K. Y. Lin and W. J. Tzeng, 
Int. J. Mod. Phys. {\bf B5}, 1913 and 2551 (1991).

\bibitem{Lin92} K. Y. Lin, J. Phys. A. {\bf 25}, 1835 (1992).

\bibitem{Moraal} H. Moraal, Physica A {\bf 203}, 91 and 103 (1994).

\bibitem{Nienhuis} B. Nienhuis,
Phys. Rev. Letts. {\bf 49}, 1062 (1982).

\bibitem{Mehlhorn} K. Mehlhorn, {\em Data Structures and Algorithms I:
Sorting and Searching}, EATCS Monographs on Theoretical Computer Science,
Springer-Verlag, Berlin (1984).

\bibitem{Polya} G. P\'{o}lya, J. Comb. Theor. {\bf 6}, 102 (1969).

\bibitem{Temperley} H. N. V. Temperley, Phys. Rev. {\bf 103}, 1 (1956).

\end{thebibliography}
\end{document}